\documentclass[conference]{IEEEtran}
\IEEEoverridecommandlockouts
\usepackage{listings}
\usepackage{comment}
\usepackage{cite}
\usepackage{amsmath,amssymb,amsfonts}
\usepackage{booktabs}
\usepackage{graphicx}
\usepackage[table,xcdraw]{xcolor}
\usepackage[normalem]{ulem}
\useunder{\uline}{\ul}{}
\usepackage{algorithmic}
\usepackage{graphicx}
\usepackage{textcomp}
\usepackage{xcolor}
\def\BibTeX{{\rm B\kern-.05em{\sc i\kern-.025em b}\kern-.08em
T\kern-.1667em\lower.7ex\hbox{E}\kern-.125emX}}

\usepackage{hyperref}
\hypersetup{
colorlinks=true,
linkcolor=blue,
filecolor=magenta,
urlcolor=cyan,
citecolor=black,
pdfpagemode=FullScreen
}

\usepackage{enumitem}

\definecolor{lightgray}{rgb}{.9,.9,.9}
\definecolor{darkgray}{rgb}{.4,.4,.4}
\definecolor{purple}{rgb}{0.65, 0.12, 0.82}

\lstdefinelanguage{script}{
  keywords={},
  keywordstyle=\color{blue}\bfseries,
  ndkeywordstyle=\color{darkgray}\bfseries,
  identifierstyle=\color{black},
  commentstyle=\color{purple}\ttfamily,
  stringstyle=\color{red}\ttfamily,
  sensitive=true,
  breaklines=true,
  escapechar=\$
}

\lstdefinelanguage{javascript}{
  keywords={do, if, in, for, let, new, try, var, case, else, enum, eval, null, this, true, void, with, await, break, catch, class, const, false, super, throw, while, yield, delete, export, import, public, return, static, switch, typeof, default, extends, finally, package, private, continue, debugger, function, arguments, interface, protected, implements, instanceof},
  keywordstyle=\color{blue}\bfseries,
  ndkeywords={class, export, boolean, throw, implements, import, this},
  ndkeywordstyle=\color{darkgray}\bfseries,
  identifierstyle=\color{black},
  sensitive=false,
  comment=[l]{//},
  morecomment=[s]{/*}{*/},
  commentstyle=\color{purple}\ttfamily,
  stringstyle=\color{red}\ttfamily,
  morestring=[b]',
  morestring=[b]",
  breaklines=true,
  escapechar=\$
}
\lstdefinelanguage{json}{
  keywords={type, data, blockSizes},
  keywordstyle=\color{purple}\bfseries,
  identifierstyle=\color{black},
  sensitive=false,
  comment=[l]{//},
  morecomment=[s]{/*}{*/},
  stringstyle=\color{red}\ttfamily,
  morestring=[b]',
  morestring=[b]",
  breaklines=true,
}

\begin{document}

\title{PASSION: Permissioned Access Control for Segmented Devices and Identity for IoT Networks
}

\author{\IEEEauthorblockN{%
Hisham Ali\IEEEauthorrefmark{1}, Mwrwan Abubakar\IEEEauthorrefmark{1}, Jawad Ahmad\IEEEauthorrefmark{1}, William J. Buchanan\IEEEauthorrefmark{1} and Zakwan Jaroucheh\IEEEauthorrefmark{1} 
}
\IEEEauthorblockA{\IEEEauthorrefmark{1} Blockpass ID Lab, Edinburgh Napier  University, Edinburgh, UK.}

}

\maketitle

\begin{abstract}

In recent years, there has been a significant proliferation of industrial Internet of Things (IoT) applications, with a wide variety of use cases being developed and put into operation. As the industrial IoT landscape expands, the establishment of secure and reliable infrastructure becomes crucial to instil trust among users and stakeholders, particularly in addressing fundamental concerns such as traceability, integrity protection, and privacy that some industries still encounter today. This paper introduces a privacy-preserving method in the industry's IoT systems using blockchain-based data access control for remote industry safety monitoring and maintaining event information confidentiality, integrity and authenticity. 
\begin{comment}
Blockchain offers different levels of data access authentication to monitor vital factors in the industry field such as temperature, barometric pressure, motion, humidity, and gases (e.g., smoke, carbon dioxide) using a secure and trusted infrastructure. Data is transmitted to clients' smartphones or computers using a Distributed Ledger. The broker serves as the server to link the messages to the desired user using Hyperledger Fabric, visualising the data on the app or computer client. 
\end{comment}
\begin{comment}
Immediate alerts allow timely action to prevent accidents, fire, and potential damage to life and property.
\end{comment}
\end{abstract}


\section{Introduction}

\begin{comment}
Safety is critical in every industry, as neglecting measures can lead to equipment damage, production interruptions, and human casualties due to harmful gas emissions. IoT plays a vital role in remotely monitoring and detecting hazardous gases, ensuring timely alerts and preventive measures for worker and environmental protection.

In recent years, the industrial Internet of Things (IoT) landscape has seen significant growth, leading to the adoption of diverse IoT applications across industries. Ensuring privacy, integrity, authenticity, and trust in IoT systems has become a paramount concern. This paper presents an advanced privacy-preserving method for industry IoT systems, integrating policy-based data access control to facilitate remote industry safety monitoring while upholding the confidentiality, integrity, and authenticity of event information. Additionally, the paper provides a comparison with decentralised systems, highlighting the strengths of Hyperledger Fabric in ensuring security and trust.
\end{comment}

Trust infrastructures within IoT networks are important in maintaining the trustworthiness of the data and of the devices used. Maintaining data privacy is challenging in centralised systems, as there is often a lack of policy-based data access control \cite{sicari2015security, wu2023blockchain}. By defining and enforcing access policies, only authorised entities can access specific data, enhancing confidentiality and safeguarding sensitive information from potential breaches. This centralised approach allows for fine-grained control over data access, enhancing privacy protection.

\begin{comment}

Hyperledger Fabric, as an enterprise-grade blockchain framework, plays a pivotal role in ensuring data integrity and authenticity. Through cryptographic techniques like digital signatures and hash functions, Hyperledger Fabric verifies the authenticity of data, detecting tampering or unauthorised modifications. The distributed ledger ensures that data remains unaltered and reliable throughout its lifecycle, enhancing the trustworthiness of the system.
\end{comment}

With blockchain solutions, we either have a permission-less approach to IoT trust, which uses a public ledger, or we can use a permission approach, such as with Hyperleder Fabric. Overall, permissioned ledgers can provide a selective access control mechanism where only trusted participants can access the distributed ledger, reducing the risk of malicious actors disrupting the system.

The aim of this paper is to propose an integrated approach for implementing access control within segmented device and identity infrastructures for IoT devices. Its core contribution is the definition and implementation of a blockchain-based access control mechanism for clear segregation between users and IoT devices, ensuring robust security, data integrity, and controlled access through policies.

  % \item Definition of a blockchain-based access control mechanism to establish clear segregation between users and IoT devices, ensuring robust security and data integrity. The proposed method incorporates policy-based data access control mechanisms. These policies enable controlled access to sensitive information and ensure that only authorised users can view and interact with specific data, enhancing overall data security.

This paper is organised as follows: Section 2 delves into current related works, providing context for our paper. Section 3 offers essential background information, covering Hyperledger Fabric, the MQTT protocol, and the publish-subscribe communication model. In Section 4, we present our system infrastructure design and its practical implementation. Section 5 conducts an evaluation, encompassing security considerations, data privacy preservation, data integrity, scalability, and throughput latency. Finally, Section 6 serves as the paper's conclusion, summarising our findings and also provides valuable suggestions for future research and development.

\begin{comment}
   : The system employs a secure and trusted infrastructure to handle data transmission. By leveraging Distributed Ledger technology (Hyperledger Fabric), the architecture ensures data integrity and reduces the risk of data tampering or unauthorised access.

   \item Introduction of a Privacy-Preserving Method: The paper introduces a privacy-preserving method specifically designed for industrial IoT systems. This method aims to ensure data privacy and confidentiality, instilling trust among users and stakeholders in the secure and reliable operation of diverse IoT applications.
\end{comment}

\section{Related Work}
\begin{comment}
The security of the Internet of Things has received significant interest from the scientific society. A significant of current researches are focused on access control and authentication in the IoT. However, the main issue with the current approaches is the centralisation. The problem with the centralised authentication approaches is that it requires to store authentication data on a centralised local server, which is prone to a single point of failure.
\end{comment}
\begin{comment}
A significant of current research is concerned with authentication and access control in IoT environments due to the smart things surge to be part of people’s daily lives on the one hand, and the amount of personal/ private information they utilize, on the other hand. However, the main issue with the current approaches is the centralisation.
\end{comment}

The problem with the centralised authentication approach is that it requires to store authentication data on a centralised local server, which is prone to a single point of failure \cite{uddin2021survey}. Similar to our work, many other approaches proposed blockchain-based authentication and access control methods  \cite{chinnasamy2021blockchain}. For instance, the work presented in \cite{chinnasamy2021blockchain} proposed a conceptual framework aimed at establishing a data-sharing system that incorporates access control mechanisms based on blockchain technology for IoT devices. The system employs three distinct smart contracts to facilitate the efficient administration of access control. These contracts include one for access control provisioning, one for authentication, and another for decision-making. Nevertheless, the implementation of a public blockchain like Ethereum in the proposed system will incur expenses for transaction processing. Similarly, the authors in \cite{nakamura2019capability} proposed a Capability-Based Access Control (CapBAC) scheme by utilising the public Ethereum blockchain technology. The authors proposed to fix some of the BlendCAC issues using a fine-grained access control model, however, no cost or performance metric was discussed. Furthermore, many studies have indicated that Role-Based Access Control (RBAC) exhibits limitations in terms of flexibility and scalability when confronted with the access control demands of IoT environments \cite{qiu2020survey}. However, there are still many researches and studies that have put forth the concept of a blockchain-based Role-Based Access Control system in different IoT domains, such as in \cite{abubakar2021decentralised} \cite{abdelrazig2021blockchain}.

Current IoT systems rely on the centralised data management model or client-server architecture to handle authentication and authorisation and to control access to IoT systems and their data. Therefore, this can put an extra cost on designing the security architecture of an IoT system due to the high costs associated with the cloud service that is needed to validate the identity of devices and applications involved in the IoT systems \cite{khan2018iot}. All information and data are gathered and shared for complete understanding, reliable delivery, and intelligent processing. This presents issues with respect to transmission costs, trust, data value, and privacy \cite{verma2022centralised}.

Blockchain is a decentralised, distributed ledger technology that has great potential to tackle the security, privacy, and scalability concerns of critical infrastructure in the IoT \cite{abubakar2022overview}. Since blockchain is considered an immutable ledger for transactions, it can track millions of IoT devices and provide highly secure communications and coordination between them \cite{shrestha2019integration}. To use blockchain technology to secure IoT devices, each device can have a unique address to send transactions. So, IoT objects don't have to trust each other because they use a consensus algorithm that lets nodes connected to the blockchain work in a trusted way. IoT devices, gateways, cloud computing, and blockchain technologies are the four layers of the decentralised architecture.

\begin{comment}
Sensing devices can be organised into a Wireless Sensor Network (WSN), which enables wireless communication and better data transfer. With the IoT concept, nodes of a WSN can connect to the internet, enabling real-time monitoring and reducing observation time. Many studies have been published in the last decade describing the applications of WSNs and IoT in various fields, including critical infrastructure.
\end{comment}

\begin{comment}
Hyperledger Fabric can be used as a secure and scalable platform for managing IoT devices and data. 
\end{comment}

\section{PASSION Permissioned Approach}

\begin{comment}
This section covers important blockchain and IoT concepts in the industry and their integration. We also discuss their role in enhancing industry safety, security, privacy, and scalability.

\subsection{Blockchain}
\end{comment}

\begin{comment}
Blockchain, introduced in 2008 by Satoshi Nakamoto, is a decentralised digital ledger technology known for its secure and tamper-proof data storage. Beyond cryptocurrencies, it finds applications in supply chain management, healthcare, digital identity verification, and voting systems, offering transparency and security.
\end{comment}

PASSION integrates Hyperledger Fabric \cite{ali2022trusted,HLF} and ensures privacy and confidentiality features, enabling participants to transact securely and share sensitive information selectively. It offers a high degree of control over the network, allowing participants to set policies, define roles, and manage access to the ledger \cite{asiri2018sybil}.

\begin{comment}
is an open-source, permissioned blockchain platform designed to support enterprise-level distributed ledger solutions for various industries such as finance, healthcare, and supply chain management. It is one of the projects under the Hyperledger umbrella, which is a collaborative effort hosted by the Linux Foundation to advance cross-industry blockchain technologies 
\end{comment}
\begin{comment}
TABLE~\ref{tab:COMPARISON} shows a comparison of different blockchain types. Hyperledger Fabric features a modular architecture that allows for flexibility and scalability, enabling users to customise their blockchain networks to meet specific business needs. 
\end{comment}

\begin{figure*}
\centering
\includegraphics[width=0.896\linewidth]{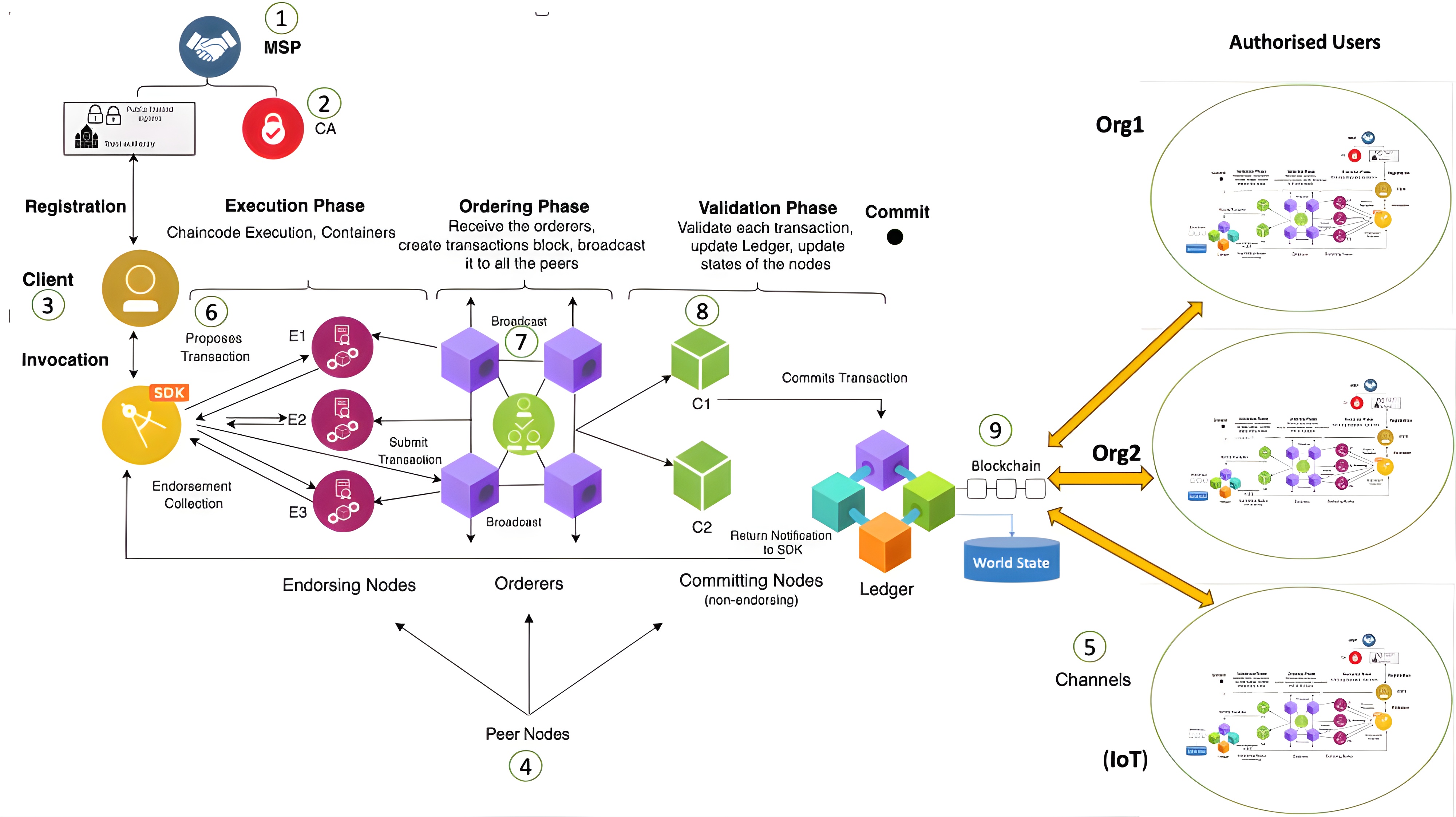}
\caption{Hyperledger Fabric-based IoT and Transaction Flow}
\label{figure:TransactionFlow}
\end{figure*}

The main elements of Hyperledger Fabric that are adopted are (Figure~\ref{figure:TransactionFlow}):

\begin{enumerate}

   \item  Membership Service Provider (MSP): The MSP is responsible for managing the certificates and identities of network participants.

    \item  Certificate Authority (CA): The CA is responsible for issuing and validating digital certificates that identify network participants.

    \item  Client: The client application interacts with the network by submitting transactions to peers and querying the ledger.

    \item  Channel: Channels are private sub-networks that allow multiple parties to transact with one another in a secure and confidential way. Channels allow a subset of nodes through the anchor node to link different Users/participants (organisations) which compose the consortium. The ledger of a channel can be accessed only by those organisations that are part of the channel. Therefore, participants can only see network features based.

    \textit{Chaincode:} Chaincode is the smart contract written in a programming language that runs on the Hyperledger Fabric network. Organisations may include different nodes connected through one channel or multiple (multi-chaincode). Notably, nodes could have multiple chain codes (engaged with multi-channel) to store the transaction data in an immutable ledger.

    \item  Peer Nodes: Peer nodes maintain copies of the ledger, execute transactions, endorse transactions, and participate in consensus.
\begin{itemize}

\item Endorser Peers (Endorsement): In (6), A transaction proposal is sent to endorsing peers, which simulate the transaction and validate its correctness according to the smart contract. If the proposal is endorsed, the endorsing peers sign it and send it back to the client.

\item Orderer Peers  (Ordering Service): The endorsed transactions are grouped into blocks and (7) sent to the ordering service. The ordering service receives endorsed transactions from peer nodes and orders them into a block (Creates a block), which is then broadcast to all peers in the network (delivers the block to each peer node). 

    \textit{Consensus:} The ordering service uses a consensus algorithm to ensure that all nodes in the network agree on the order of the blocks. This ensures that the ledger is consistent across all nodes in the network.

     \item Validation Peers: In (8), Peers validate the block and transactions contained within it, checking the digital signatures and endorsement policies. If the block is valid, the transactions are committed to the ledger.

    \item Update the ledger (9).

     \end{itemize} 

\end{enumerate}

\begin{comment}
Overall, Hyperledger Fabric is designed to enable secure, transparent, and efficient interactions between different organisations, facilitating trust and collaboration in complex business ecosystems.
\end{comment}

\subsection{Hyperledger Fabric-Based IoT Architecture components}
The architecture for Hyperledger Fabric IoT includes the following components:

\begin{itemize}

 \item \textit{Applications:} The applications layer includes the software and services that consume and process the data stored on the blockchain. This can include data analytics, machine learning algorithms, and other applications that provide insights or actions based on the analysis of the data.
 \item \textit{Application Support Layer:} The data processing layer plays a critical role in IoT systems as it enables the extraction of meaningful insights from the large amounts of data generated by connected devices. The layer typically consists of several components, including Data collection; Data filtering and preprocessing; Data storage; Data analytics; and Data visualisation and reporting.

 \item \textit{IoT devices/sensors:} This layer consists of the physical devices and sensors that collect data from the environment.

 \item \textit{Network Layer:} The network layer, also known as the transmission layer, is responsible for managing network connectivity-related tasks such as authentication, authorisation, accountability, and IoT transport management data. It acts as a bridge between the perception and application layers, transmitting the data collected from physical objects. The transmission medium can be either wireless or wired, and it is responsible for connecting smart devices, network devices, and networks together. Moreover, the following are such examples:

\begin{itemize}

 \item \textit{Gateway:} The gateway acts as a bridge between the IoT devices and the blockchain network and is responsible for collecting data from the devices and sending it to the blockchain.

 \item \textit{Blockchain network:} This layer includes the distributed network of nodes that form the Hyperledger Fabric blockchain. The nodes are responsible for processing transactions, validating data, and reaching consensus.

 \item \textit{Smart contracts: }Smart contracts are the business logic that defines how the data from the IoT devices is processed, stored, and shared on the blockchain. They can be used to define data schemas, process data, and enforce rules for accessing and sharing data.

\end{itemize}
 \end{itemize}

 \begin{comment}
 \begin{figure}
\centering
\includegraphics[width=0.8\linewidth]{Figures/Layers.png}
\caption{Industry safety monitoring and IoT system architecture using Hyperledger Fabric}
\label{figure:TransactionFlow}
\end{figure}
\end{comment}

   \begin{figure}
\centering
\includegraphics[width=1\linewidth]{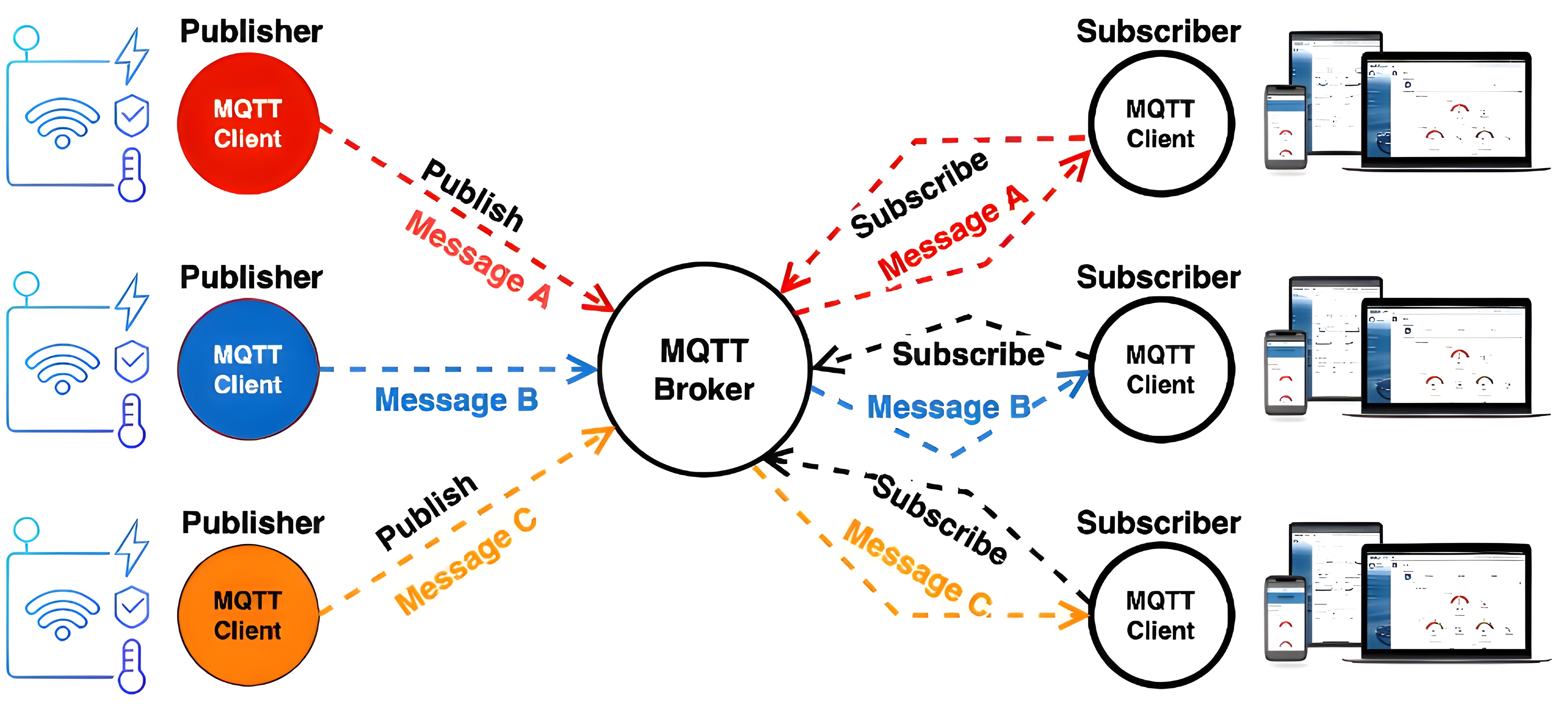}
\caption{The publish-subscribe IoT communication model with an MQTT Broker}
\label{figure:Thepublish}
\end{figure}

\subsection{MQTT}

MQTT (Message Queuing Telemetry Transport) is a lightweight and efficient messaging protocol designed for low-bandwidth, unreliable networks \cite{boccadoro2019quakesense}. It enables communication between devices and applications in the Internet of Things (IoT) and other scenarios where a simple and efficient messaging system is needed.

\textit{Key features of MQTT include:}

\begin{itemize}

    \item Publish-Subscribe Model: MQTT uses a publish-subscribe model, where devices can publish messages to topics, and other devices (subscribers) can receive those messages by subscribing to specific topics.

    \item QoS Levels: MQTT supports three Quality of Service (QoS) levels to ensure message delivery reliability:

    \begin{itemize}
    
   \item \textit{QoS 0:} At most once - Fire and forget; no acknowledgment is sent.

   \item \textit{QoS 1:} At least once - Messages are guaranteed to be delivered, but duplicates may occur.

   \item \textit{QoS 2:} Exactly once - Messages are ensured to be delivered only once and in the correct order.
   
    \end{itemize}
    
    \item Lightweight: MQTT is designed to be efficient and lightweight, making it suitable for resource-constrained devices and low-bandwidth networks.

    \item Persistent Session: MQTT supports persistent sessions, allowing subscribers to receive messages sent while they were offline when they reconnect.

    \item Retained Messages: Publishers can set messages as "retained," meaning that the last published message on a topic will be saved and delivered to new subscribers when they connect.

    \item Scalability: MQTT is scalable and can support a large number of clients, making it suitable for various IoT and real-time messaging applications.

\end{itemize}    

MQTT has become widely used in IoT applications due to its simplicity, efficiency, and ability to handle unreliable network conditions. It has several implementations and is supported by many platforms and programming languages.

\begin{figure}[ht!]
\centering
\includegraphics[width=1\linewidth]{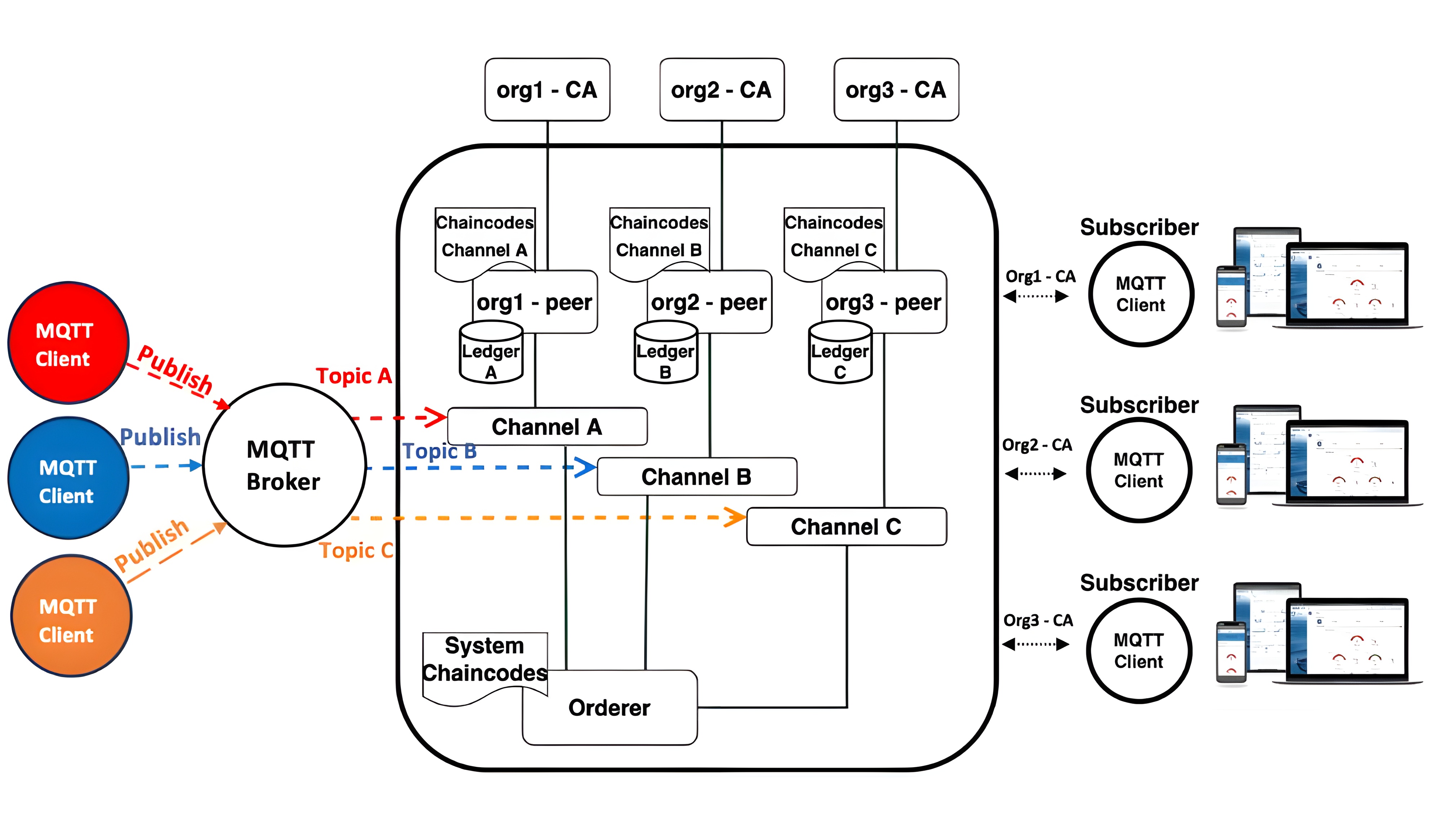}
\caption{Industrial data security with an MQTT Broker, using Hyperledger Fabric, maintaining data integrity and privacy.}
\label{figure:Hyperledger}
\end{figure}

\subsection{The publish-subscribe communication model}

The publish-subscribe communication model consists of three main components: publishers, subscribers, and a message broker (or messaging system).

\textit{Publishers} are responsible for generating and sending messages to the messaging system, and they do so without having knowledge of the \textit{subscribers}' identities \cite{palmieri2019mqttsa}. Instead, publishers publish messages on specific topics. On the other hand, subscribers express their interest in receiving messages related to particular topics and register themselves with the messaging system accordingly. When a publisher sends a message to a topic, the message \textit{broker} acts as an intermediary. It receives the message and ensures that all registered subscribers interested in that topic receive the message. The message broker facilitates communication between publishers and subscribers without requiring direct interaction between them. This decoupling allows for a flexible and scalable communication approach, making the publish-subscribe model well-suited for various applications, including real-time data streaming, Internet of Things (IoT) systems, financial services, and social media platforms.

In PASSION, Hyperledger Fabric clients play the roles of subscribers and publishers, utilising public and private keys for secure interactions with the MQTT broker. Subscribers connect to the broker to subscribe to specific topics and receive data from publishers, which are resource-constrained IoT devices like sensors. Publishers authenticate with the MQTT broker to publish their sensor readings on designated topics. When establishing a connection, both subscribers and publishers receive a challenge from the smart contract, ensuring only authorised addresses receive it. Using their private keys, they cryptographically prove their identity by signing the challenge and interacting with the smart contract. This challenge acts as a one-time password, providing secure access to the broker. By integrating Hyperledger Fabric, our system ensures a reliable and secure communication model suited for IoT and real-time data applications.

\section{Implementation}

In this section, we showcase the proposed infrastructure employed in our study, including the setups and decision-making processes. We also delve into the specifics of the implementation that influenced our experiments on the network model. We have carried out a real-life application, which entails utilising IoT sensors to acquire environmental data. 

\subsubsection{Environment Setup and Test}

 This section includes network performance and functionality tests. A network has been set up to demonstrate the integration of blockchain and IoT, comprising three organisations: Org1, Org2 and Org3 (IoT). The application client receives camera streaming data, temperature, humidity, and gas data from the MQTT broker and updates the blockchain network accordingly. The organisation's TLS-CA server offers TLS (Transport Layer Security) to all blockchain nodes, including the ordering and CA (Certification Authority) servers and users, to secure the network connection. In addition, X509 certificates are issued to all the members and actors in the organisation's blockchain network by the CA server. Then, the benchmark engine interacts with chain code to deploy, run, analyse, and generate network performance reports. Table~\ref{tab:network nodes} defines the network nodes present in our experimental network. Overall, Hyperledger Fabric is run within a Docker container on a Ubuntu instance.

\begin{table}[]
\caption{The IoT nodes present in our experimental network}
\label{tab:network nodes}
\resizebox{\columnwidth}{!}{%
\begin{tabular}{|l|l|l|l|}
\hline
\rowcolor[HTML]{C0C0C0} 
\textbf{Client} & \textbf{IP Adress} & \textbf{Operation System}                               & \textbf{Fabric Version}       \\ \hline
\rowcolor[HTML]{EFEFEF} 
Org1-TLS-CA    & 10.0.1.10 & Ubuntu 20.04 Linux 2.4 (64-bit) & 2.3.0 \\ \hline
\rowcolor[HTML]{EFEFEF} 
Org1-CA        & 10.0.1.20 & Ubuntu 20.04 Linux 2.4 (64-bit) & 2.3.0 \\ \hline
\rowcolor[HTML]{EFEFEF} 
peer@org1      & 10.0.1.30 & Ubuntu 20.04 Linux 2.4 (64-bit) & 2.3.0 \\ \hline
\rowcolor[HTML]{EFEFEF} 
Org2-TLS-CA    & 10.0.2.10 & Ubuntu 20.04 Linux 2.4 (64-bit) & 2.3.0 \\ \hline
\rowcolor[HTML]{EFEFEF} 
Org2-CA        & 10.0.2.20 & Ubuntu 20.04 Linux 2.4 (64-bit) & 2.3.0 \\ \hline
\rowcolor[HTML]{EFEFEF} 
peer@org2      & 10.0.2.30 & Ubuntu 20.04 Linux 2.4 (64-bit) & 2.3.0 \\ \hline
\rowcolor[HTML]{EFEFEF} 
IoT-TLS-CA      & 10.0.4.10          & \cellcolor[HTML]{EFEFEF}Ubuntu 20.04 Linux 2.4 (64-bit) & \cellcolor[HTML]{EFEFEF}2.3.0 \\ \hline
\rowcolor[HTML]{EFEFEF} 
IoT-CA          & 10.0.4.20          & \cellcolor[HTML]{EFEFEF}Ubuntu 20.04 Linux 2.4 (64-bit) & \cellcolor[HTML]{EFEFEF}2.3.0 \\ \hline
\rowcolor[HTML]{EFEFEF} 
peer@IoT        & 192.168.10.30      & \cellcolor[HTML]{EFEFEF}Ubuntu 20.04 Linux 2.4 (64-bit) & \cellcolor[HTML]{EFEFEF}2.3.0 \\ \hline
\rowcolor[HTML]{EFEFEF} 
Orderer-TLS-CA & 10.0.5.10 & Ubuntu 20.04 Linux 2.4 (64-bit) & 2.3.0 \\ \hline
\rowcolor[HTML]{EFEFEF} 
Orderer-CA     & 10.0.5.20 & Ubuntu 20.04 Linux 2.4 (64-bit) & 2.3.0 \\ \hline
\rowcolor[HTML]{EFEFEF} 
Solo@Orderer   & 10.0.5.30 & Ubuntu 20.04 Linux 2.4 (64-bit) & 2.3.0 \\ \hline
\end{tabular}%
}
\end{table}

\begin{comment}
Hyperledger Fabric network setup has prerequisites, with: (1) Fabric binaries package version 2.2.0 installed on (2) Ubuntu version 20.04. (3) Docker, version 20.10.7, and (4) Docker Compose, version 1.25.0, are used to generate and operate the entities in the network.  
\end{comment}

\begin{comment}
Most notably, Hyperledger Fabric, by default, comes with three different optional languages: Golang, Java, and JavaScript. We built chaincode (smart contract) using JavaScript (Node.js version 10.19.0).
\end{comment}

% Please add the following required packages to your document preamble:
% \\usepackage[table,xcdraw]{xcolor}
% If you use beamer only pass "xcolor=table" option, i.e. \documentclass[xcolor=table]{beamer}
\begin{comment}
\begin{table}[ht!]
\centering
\caption{Hyperledger Fabric environment setup}
\label{tab:my-table3}
\begin{tabular}{|l|l|}
\hline
\rowcolor[HTML]{9B9B9B} 
{System \& Tools} & {Version} \\ \hline
\rowcolor[HTML]{EFEFEF} 
CPU         & 2.6 GHz Quad-Core i7   \\ \hline

\rowcolor[HTML]{EFEFEF} 
Memory      & 16 GB 1600 MHz DDR3    \\ \hline

\rowcolor[HTML]{EFEFEF} 
Hard disk        & 1T    \\ \hline

\rowcolor[HTML]{EFEFEF} 
Operating System         & Ubuntu 20.04     \\ \hline
\rowcolor[HTML]{EFEFEF} 
Hyperledger Fabric       & 2.3.0            \\ \hline
\rowcolor[HTML]{EFEFEF} 
Docker                   & 20.10.7          \\ \hline
\rowcolor[HTML]{EFEFEF} 
Docker-compose           & 1.25.0           \\ \hline
\rowcolor[HTML]{EFEFEF} 
Node.js                  & 10.19.0          \\ \hline
\end{tabular}
\end{table}
\end{comment}

\begin{figure*}
\centering
\includegraphics[width=1\linewidth]{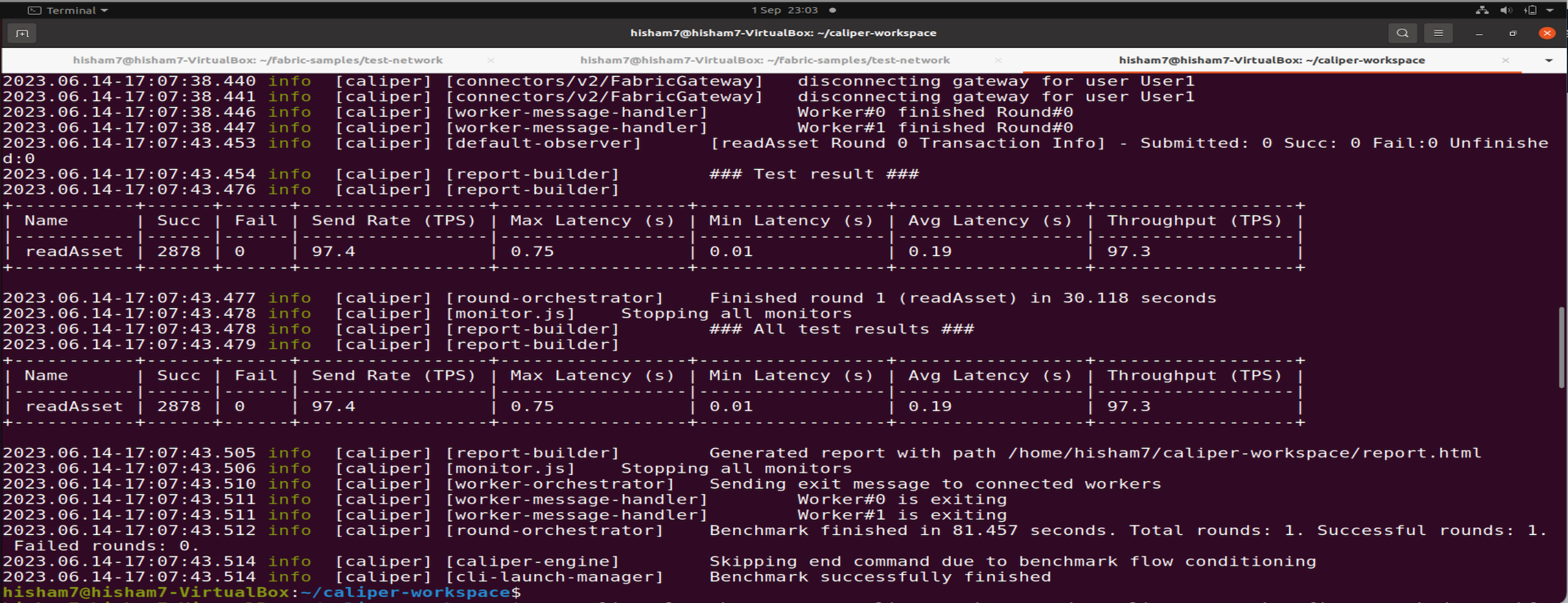}
\caption{Running the Caliper benchmark and obtaining the performance report for the IoT network}
\label{figure:runcaliper}
\end{figure*}

\subsubsection{Interaction System Implementation}

The implementation extends the asset-transferbasic/chain code-javascript using a \textit{Solo Ordering Service}, provided by the Hyperledger Fabric and test network. This performance tests the smart contract on a Fabric network using Caliper. The basic workflow of this whole system is:

\textit{Implementation of smart contracts:} We used Hyperledger Fabric smart contracts as a proof of concept (chaincode). The Hyperledger Fabric network was selected because of the features mentioned that related to its unique design as modular and extensible, delivering confidentiality, scalability and privacy. We employ the JavaScript programming language to deploy smart contracts in Hyperledger Fabric systems. Finally, we test it by installing it, approving its definition, committing it to the channel, and invoking the chain code. By executing the given chain code, user interaction within the Hyperledger Fabric ledger is feasible. Notably, One node can have multiple smart contracts, making it possible to monitor different sensor readings from this node (the user).

The chain code is in charge of dealing with various data queries. As a result, the system implementation begins by defining certain chaincode operations, such as querying and retrieving data lineage. The chain code allows the authorised user to get a URL using his identity to obtain sensor information. In other words, sensor information in the ledger is only accessible to users who have been given permission to use it since the chain code can only give out authority after the user has been verified. The chaincode is intended to facilitate various data and traceability processes inside the ledger and attached-chain storage. The proposed system’s chain code-specific operations include storing data on an item’s world state, querying item checksums, retrieving an object with the relevant transaction ID, extracting the version of an object based on its transaction ID, retrieving the lineage of the data item, retrieving the history of a data object, querying the key-range of the list of items (AssetsID), retrieving the specific sensor's information, and providing a specific version of an object. Assets represent the variable value of items that may be exchanged on blockchain platforms during transaction execution. The implemented system is made up of distributed peer nodes that serve as the hub for communication among network parts, as shown in Figure~\ref{figure:TransactionFlow}. The suggested model’s performance was tested in terms of system throughput, send rate, latency, and resource usage (memory, CPU, network). The scope of the investigation was expanded to examine the latency and scalability of different transaction loads, transaction duration TPS, and Asset batch sizes.

The benchmark involves evaluating ’getAssetsFrom- Batch’ gateway transactions for the fixed-asset smart contract; the endorsement policy and the network are implemented within LevelDB and CouchDB state databases. Fabric supports two alternatives for a key-value store, CouchDB and LevelDB, to maintain the current state. Both are key-value stores; while LevelDB is an embedded database, CouchDB uses a client-server model (accessed using a REST API over a secure HTTP) and supports a document/JSON data model. Each transaction obtains a collection of assets from the world state database, which is comprised of a random selection of available UUIDs (Universal Unique identifiers). Following rounds, increase the batch size of assets acquired from the world state database with a fixed load.

The measurements were carried out using a command-line interface (CLI) by configuring the Caliper benchmarking tool using the benchmark workspace, network module, and workload to monitor the system’s performance. The test was carried out by simulating a specific transaction load through Org1 “User A” and Org2 “User B” and IoT (Org3). The edge server saved the identities of all connected nodes and authenticated them inside a trustworthy Hyperledger Fabric environment by applying the mutual authentication mechanism. The suggested model’s performance was evaluated for a variety of workloads and environmental conditions. Furthermore, a diverse set of interaction performances was observed to investigate the improvement or deterioration induced by different model parameters and setups. We evaluated Hyperledger Fabric V2.3.0 benchmarking, real-time data reporting, and resource consumption statistics were gathered and monitored. The following steps provide examples of various functions through Hyperledger infrastructure configuration and network performance benchmarking: 1. Bring up the test network and create the channel; 2. Package and install the smart contract; 3. Approve a chain code definition; 4. Commit the chain code definition to the channel; 5. Invoke the chaincode; and 6. Run Caliper Benchmark and get the network performance report by monitoring IoT network latency, send rate and throughput as shown in Figure~\ref{figure:runcaliper}

\section{Evaluation}

Hyperledger Fabric IoT architecture can provide several benefits, such as improved security, scalability, and transparency. It can also enable the development of new business models and services that leverage the data collected from IoT devices. However, implementing this architecture requires a deep understanding of both blockchain and IoT technologies, as well as experience in integrating and deploying these technologies in a real-world setting.

IoT devices generate various types of data, most of which is unstructured. For instance, cameras capture images and videos, microphones record external sounds, and sensors detect physical signals like gas, temperature and humidity, all of which are converted into digital data. These real-time data cannot be directly stored in relational databases; they must be pushed promptly to authorised users. Voice and video data are streamed, encoded, and sent to the cloud server via WiFi or 4G, resulting in the generation of a resource URL that users can use to access the data. On the other hand, for sensor data, the device sends the data to a topic through an MQTT-based service or other protocols, and the server pushes the message to the client after authorising it through the Hyperledger Fabric network. This kind of data is mainly used to control IoT devices and perform various operations. Clients can send requests to the server through a restful API based on HTTP(s), and the server can send control signals back to the device through MQTT or other protocols:

\begin{comment}
\subsection{Security Consideration}

Security is a top priority when using blockchain in IoT. It is especially crucial in industries and IoT setups with many connected devices.

Our method for access control and authentication is vital for data and network security. It helps us identify devices, protect against compromised nodes, and control data access levels.

We have chosen a consortium approach, like Hyperledger Fabric, for access control. This means organisations collaborate to set rules for the blockchain network.

Fabric Policies are key. They enforce access control and set resource access limits. Fine-grained control ensures only authorised users can access certain functions, adding extra security. Endorsement policies specify who must approve transactions, ensuring only valid ones go on the blockchain.

Implementing these measures improves IoT devices and blockchain network security. Keeping consortium rules updated and regularly reviewing access control policies is essential, as security is an ongoing process.
\end{comment}

\begin{itemize}
    \item Data Privacy-preserving. Data privacy and data trading are crucial in today's digital landscape. Blockchain technologies, such as Hyperledger Fabric, enhance data privacy by securely storing and sharing information in a decentralised and immutable manner. Access control features in Hyperledger Fabric ensure that only authorised parties can access and modify data and smart contracts. In our proposed model, Hyperledger Fabric channels facilitate secure data trading between broker users with granular access control, granting specific permissions to individual users or organisations. This empowers organisations to protect sensitive information and securely trade data.
    \item Data integrity.  Ensuring data integrity is crucial for industrial safety, with goals of assurance, completeness, consistency, and dependability throughout the data life cycle. Blockchain's decentralised system, using hashed blocks, guarantees data integrity and mitigates challenges from cloud services connected to IoT devices. Configuring IoT devices as direct blockchain nodes enhances data reliability, removing human intervention and external system reliance. This approach reinforces the integrity of industrial IoT data, promoting a secure and trustworthy environment.
    \item Scalability, Throughput and Latency. Hyperledger Fabric's Execute-Order-Validate method, which separates transaction execution and ordering, is a key advantage. This separation boosts scalability, enhances performance, and reduces node workload. Unlike other blockchain designs, Fabric's approach introduces parallel transaction processing, addressing smart contract non-determinism. This results in higher throughput and lower latency, creating an efficient and high-performance blockchain ecosystem. In summary, this approach promotes privacy, trust, scalability, and access control in secure IoT data systems, setting the stage for secure information exchange while maintaining privacy and trust.
\end{itemize}

\section{Conclusion}

The PASSION approach defines permissioned approach to access control for segmented Devices and identity for IoT networks and allows an entity to carefully control the usage of devices and data within a trusted infrastructure. As part of our future work, we plan to expand the IoT network, evaluate hardware capabilities, and explore innovative consensus methods and scalability for improved data processing and responsiveness during safety-critical events.

%In conclusion, our blockchain solution enhances data security, privacy, and reliability in industrial IoT, allowing organisations to embrace IoT technology while maintaining worker and facility safety and privacy.

\bibliographystyle{IEEEtran}
\bibliography{ref}

\end{document}